\documentclass[aip,rsi, amsmath,amssymb, reprint,]{revtex4-1}
\pdfoutput=1
\usepackage{graphicx}
\usepackage{dcolumn}
\usepackage{bm}

\usepackage{pgfplots}
\usepackage{pgfplotstable}
\usepgfplotslibrary{external} 
\usepgfplotslibrary{fillbetween}
\usepgfplotslibrary{colorbrewer}

\usetikzlibrary{external}
\usepackage{subfigure}
\usepackage{cleveref}

\pgfmathsetmacro{\Tfreeze}{364}
\pgfmathsetmacro{\Tjump}{442.335666891}
\pgfmathsetmacro{\tfreeze}{0.483233608052}
\pgfmathsetmacro{\tjump}{0.535762818744}
\pgfmathsetmacro{\Tc}{537}

\definecolor{android_blue}{RGB}{51,181,229}
\definecolor{android_dark_blue}{RGB}{0,153,204}
\definecolor{android_pink}{RGB}{170,102,204}
\definecolor{android_purple}{RGB}{156,39,176}
\definecolor{android_dark_pink}{RGB}{153,51,204}
\definecolor{android_green}{RGB}{153,204,0}
\definecolor{android_dark_green}{RGB}{102,153,0}
\definecolor{android_orange}{RGB}{255,152,0}
\definecolor{android_dark_orange}{RGB}{255,152,0}
\definecolor{android_red}{RGB}{255,68,68}
\definecolor{android_dark_red}{RGB}{204,0,0}
\definecolor{android_pink}{RGB}{156,39,176}
\definecolor{android_grey}{RGB}{158,158,158}

\pgfplotsset{grid style={dashed,grey,opacity=0.5}}

\pgfplotscreateplotcyclelist{peak_temp}{
{color=android_dark_blue,line width=1.0pt,mark=x,mark size=2pt,mark options={line width=1.0pt},line join=round},
{color=android_red,line width=1.0pt,mark=o,mark size=2pt,mark options={line width=1.0pt},line join=round},
{color=android_dark_green,line width=0.5pt,mark=|,mark size=2pt,mark options={line width=0.75pt},line join=round},
{color=black,line width=0.75pt,mark size=2pt,mark=triangle,mark options={line width=0.75pt},line join=round},
{color=android_blue,line width=0.5pt,mark=square,mark size=2pt,mark options={line width=0.75pt},line join=round},
	  {color=android_pink,line width=0.5pt,mark=diamond,mark size=2pt,mark options={line width=0.75pt},line join=round},
	  {color=android_orange,line width=0.5pt,mark size=2pt,mark options={line width=0.75pt},line join=round}
	  }

\begin{document}
\pgfplotsset{colormap/RdBu-9}

\preprint{AIP/123-QED}

\title{Noise reduction in heat-assisted magnetic recording by optimizing a high/low Tc bilayer structure}

\author{O. Muthsam}
 \email{olivia.muthsam@univie.ac.at}
\author{C. Vogler}%
\author{D. Suess}
 \affiliation{ 
University of Vienna, Physics of Functional Materials
}%

\date{\today}

\begin{abstract}
It is assumed that heat-assisted magnetic recording (HAMR) is the recording technique of the future. For pure hard magnetic grains in high density media with an average diameter of $5$\,nm and a height of $10$\,nm the switching probability is not sufficiently high for the use in bit-patterned media. Using a bilayer structure with 50$\%$ hard magnetic material with low Curie temperature and 50$\%$ soft magnetic material with high Curie temperature to obtain more than 99.2$\%$ switching probability, leads to very large jitter. We propose an optimized material composition to reach a switching probability of $P_{\mathrm{switch}}>99.2\%$ and simultaneously achieve the narrow transition jitter of pure hard magnetic material. Simulations with a continuous laser spot were performed with the atomistic simulation program VAMPIRE for a single cylindrical recording grain with a diameter of 5\,nm and a height of 10\,nm. Different configurations of soft magnetic material and different amounts of hard and soft magnetic material were tested and discussed. Within our analysis, a composition with 20$\%$ soft magnetic and $80\%$ hard magnetic material reaches the best results with a switching probability $P_{\mathrm{switch}}>99.2\%$, an off-track jitter parameter $\sigma_{\mathrm{off},80/20}=14.2$\,K and a down-track jitter parameter $\sigma_{\mathrm{down},80/20}=0.49$\,nm.
\end{abstract}

\maketitle

\section{\label{sec:level1}INTRODUCTION}

Heat-assisted magnetic recording (HAMR) \cite{ersteshamr,fan,hamr1,hamr} is considered to be a promising approach to increase the areal storage density of recording media in the future. High areal density means small recording grains which require high anisotropy to be thermally stable. The available field of the write head limits the anisotropy of the grain. HAMR overcomes this so-called recording trilemma by using a local heat pulse to heat the material near or above the Curie temperature. By doing this, the coercivity of the material is reduced such that the available head field is sufficient to switch the grain. However, thermally written-in errors are a serious problem of HAMR. It has been shown that for pure FePt-like hard-magnetic grains with a height of 10\,nm and a diameter of 5\,nm, the switching probability of one grain is clearly below 99$\%$ \cite{suess} which is too low for practical use in bit-patterned media. The idea to overcome this problem is to use a bilayer structure with graded Curie temperature which consists of a hard magnetic layer with low Curie temperature and a soft magnetic layer with high Curie temperature \cite{suess1}. Similar to this, a thermal spring magnetic medium was also proposed by Coffey et al \cite{coffey}. Nevertheless, it was shown \cite{areal} that using a 50/50 low/high Tc bilayer structure leads to a significant increase of both the down-track and the off-track jitter parameter. In this paper, we optimize an exchange coupled grain to obtain a switching probability of $P_{\mathrm{switch}}>99.2\%$ while maintaining the low down-track and off-track jitter of a pure FePt grain with the same dimensions. This is achieved by varying the composition of the soft magnetic layer as well as the ratio between the soft and the hard magnetic layer. The simulations were performed with the atomistic simulation program VAMPIRE, which solves the stochastic Landau-Lifshitz-Gilbert equation \cite{evans}.\\
The structure of this work is as follows: In Section II, the HAMR models that are used in the simulations as well as the composition of the materials and the optimization parameters are explained. Section III summarizes the results and shows which material composition works best. In Section IV, the results are discussed.

\section{\label{sec:level2}Modeling HAMR}

For the simulations a cylindrical recording grain is considered with a height of 10\,nm and a diameter of 5\,nm. It can be interpreted as one island of a patterned media design with ultra high density. A simple cubic crystal structure is used. In the atomistic simulations, only nearest neighbor exchange interactions between the atoms are included. A continuous laser pulse with Gaussian shape and full width at half maximum (FWHM) of 20\,nm is assumed in all simulations. The temperature profile of the heat pulse is given by

\begin{align}
T(x,y,t)= (T_{\mathrm{write}}-T_{\mathrm{min}})e^{-\frac{(x-vt)^2+y^2}{2\sigma^2}} + T_{\mathrm{min}}
\end{align}

with

\begin{align}
\sigma=\frac{\mathrm{FWHM}}{\sqrt{8\ln(2)}}
\end{align}

and

\begin{align}
T_{\mathrm{peak}}=(T_{\mathrm{write}}-T_{\mathrm{min}})e^{-\frac{y^2}{2\sigma^2}}+T_{\mathrm{min}}.
\label{equation}
\end{align}

The speed $v$ of the write head is assumed to be 20\,m/s. $x_0=vt$ denotes the down-track position of the write head with respect to the center of the bit. $x$ and $y$ label the down-track and the off-track position of the grain, respectively. In our simulations both the down-track position $x$ and the off-track position $y$ are variable. The final temperature of all simulations is $T_{\mathrm{min}}=270$\,K. Since the laser pulse is continuously switched on, the correct timing of the field pulse is very important. The applied field is modeled as a trapezoidal field with a write frequency of 1\,Ghz and a field rise and decay time of 0.1\,ns. The field strength is assumed to be +0.8\,T and -0.8\,T in $z$-direction. Initially, the magnetization of each grain points in $+z$-direction. The trapezoidal field tries to switch the magnetization of the grain from $+z$-direction to $-z$-direction. At the end of every simulation, it is evaluated if the bit has switched or not.

\subsection{HARD MAGNETIC LAYER}
The composition of the FePt like hard magnetic material is the same in all simulations. Only the amount of the hard magnetic material is optimized later. The parameters are chosen as follows: For the damping constant we use $\alpha_{\mathrm{HM}}=0.1.$ The atomistic spin moment in $\mu_{\mathrm{HM}}=1.7\,\mu_{\mathrm{B}}$ which corresponds to a saturation polarization $J_{\mathrm{HM}}=1.42$\,T. The exchange energy within the hard magnetic material is $J_{ij,\mathrm{HM}}=5.18 \times 10^{-21}$\,J/link. We use uniaxial anisotropy in $z$-direction with an anisotropy constant $9.12 \times 10^{-23}$\,J/link which corresponds to an uniaxial anisotropy $K_{1,\mathrm{HM}}=6.6$\,MJ/$m^3$. This material composition was also used in former simulations by Suess \textit{et al} \cite{suess} and Vogler \textit{et al}\cite{areal}.

\subsection{OPTIONAL PARAMETERS}

\begin{figure}
\centering
\includegraphics[width=0.6\linewidth]{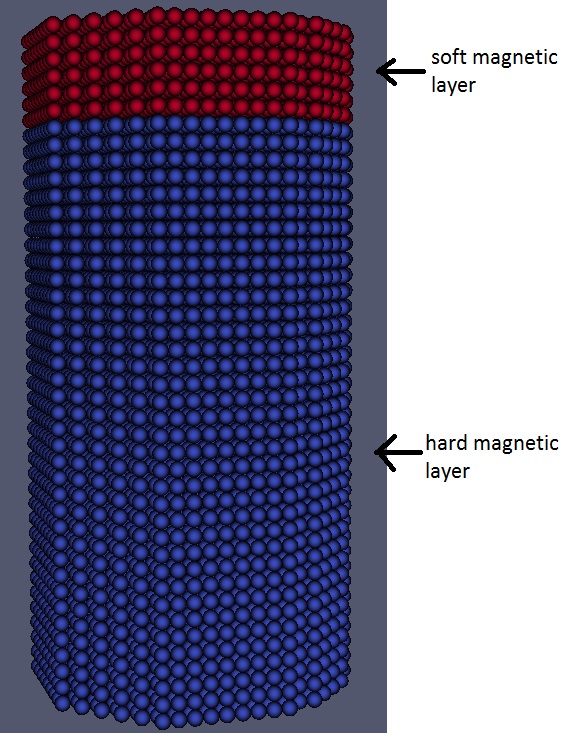}
\caption{Schematic representation of the atoms within the cylindrical recording grain with a bilayer structure composed of soft magnetic material (red) and  hard magnetic material (blue). }
\label{8020material}
\end{figure}

In order to find the material composition which maximizes the switching probability and simultaneously minimizes the transition jitter, different parameters of the soft magnetic material are varied and simulations for the different material compositions are performed. How\-ever, some parameters of the soft magnetic material are specifically chosen to model the material as realistic as possible. The fixed parameters, on one hand, are the damping constant $\alpha_{\mathrm{SM}}=0.1$ and the anisotropy constant $K_{1,\mathrm{SM}}$ which is set to zero. On the other hand, the exchange energy within the layer, the exchange energy between the layers and the atomistic spin moment are variable. The amount of hard magnetic material in these simulations is always 50$\%$.\\
First, the optimal parameters for the soft magnetic material are determined, then the amounts of soft and hard magnetic material are additionally varied to further optimize the recording grains.
\Cref{8020material} shows such a bilayer composition with 20$\%$ soft magnetic material and 80$\%$ hard magnetic material.

\section{RESULTS}
\subsection{HARD MAGNETIC LAYER}

\begin{figure}
\centering
\includegraphics[width=1.0\linewidth]{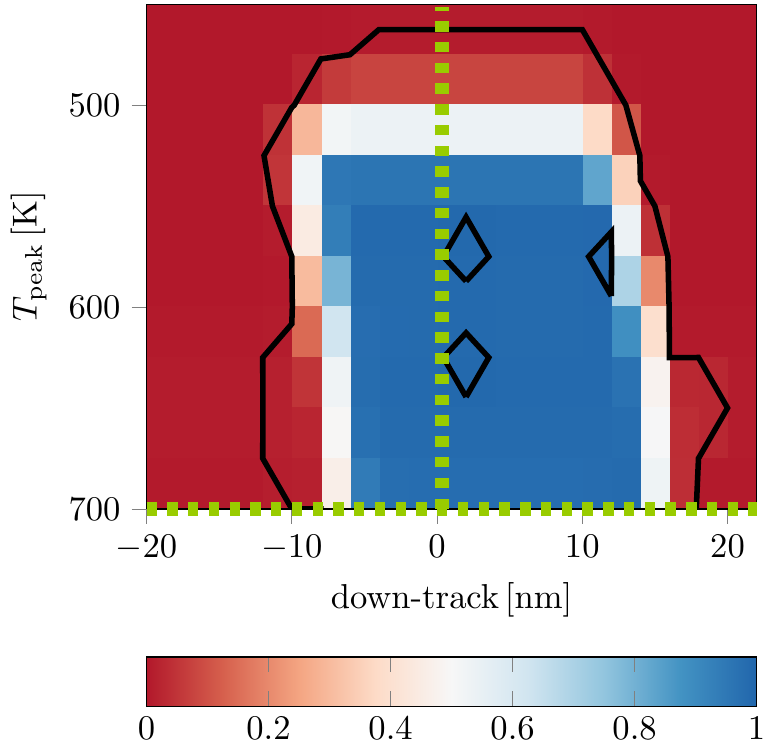}
  \caption{Switching probability phase diagram of a pure FePt like hard magnetic grain. The contour lines indicate the transition between areas with switching probability less than 1$\%$ (red) and areas with switching probability higher than 99.2$\%$ (blue). The dashed lines mark the switching probability curves of \Cref{jitter}. }
  \label{feptphase}
\end{figure}

\begin{figure}
\centering
\includegraphics[width=1.0\linewidth]{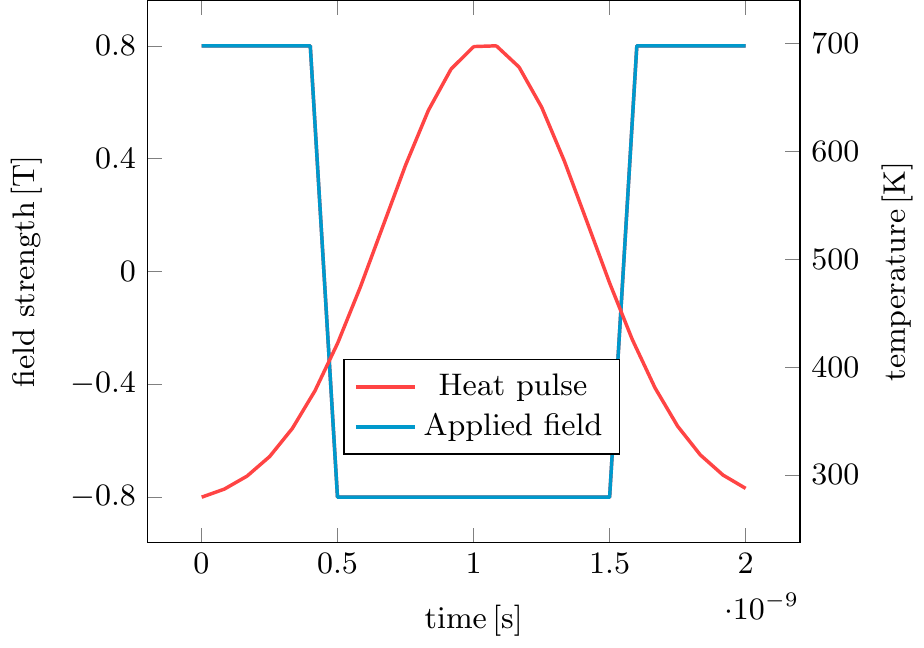}
\caption{Schematic representation of a trapezoidal field and a Gaussian heat pulse for the HAMR simulations at down-track position $x=0$\,nm.}
\label{feldmitpulse}
\end{figure}

First, a phase diagram, where the switching probability depending on the the down-track position $x$ and the off-track position $y$ is computed for a pure FePt like grain, see \Cref{feptphase}. If the write temperature $T_{\mathrm{write}}$ of the heat spot is fixed, every peak temperature $T_{\mathrm{peak}}$ corresponds to an unique off-track position $y$ (see \cref{equation}). Therefore, in the phase diagram, the switching probability is shown as a function of the down-track position $x$ and the, to $y$ corresponding, peak temperature $T_{\mathrm{peak}}$. \\
The schematic position between the heat pulse and the trapezoidal field at down-track position $x=0$\,nm can be seen in \Cref{feldmitpulse}. In the simulations, only the cooling of the heat pulse is considered, i.e. the simulations do not start before the peak of the heat pulse.\\
The resolution in the down-track direction is $\Delta x=2$\,nm and the resolution in the peak temperature direction is $\Delta T_{\mathrm{peak}}=25$\,K. The velocity of the write head is assumed to be $v_{\mathrm{h}}=20$\,m/s. In each phase point, 128 trajectories are simulated. Thus, the switching probability phase diagram contains almost 60.000 switching trajectories with a length of 2\,ns. The areas with less than 1$\%$ switching and the areas with more than $99.2\%$ switching are marked by the contour lines. The phase diagram of the pure hard magnetic grain shows only a few areas with complete switching. In particular, no complete switching occurs for high peak temperatures larger than 650\,K. This shows the high DC noise of pure hard magnetic grains.

\begin{figure}
\centering
\subfigure{\includegraphics[width=\linewidth]{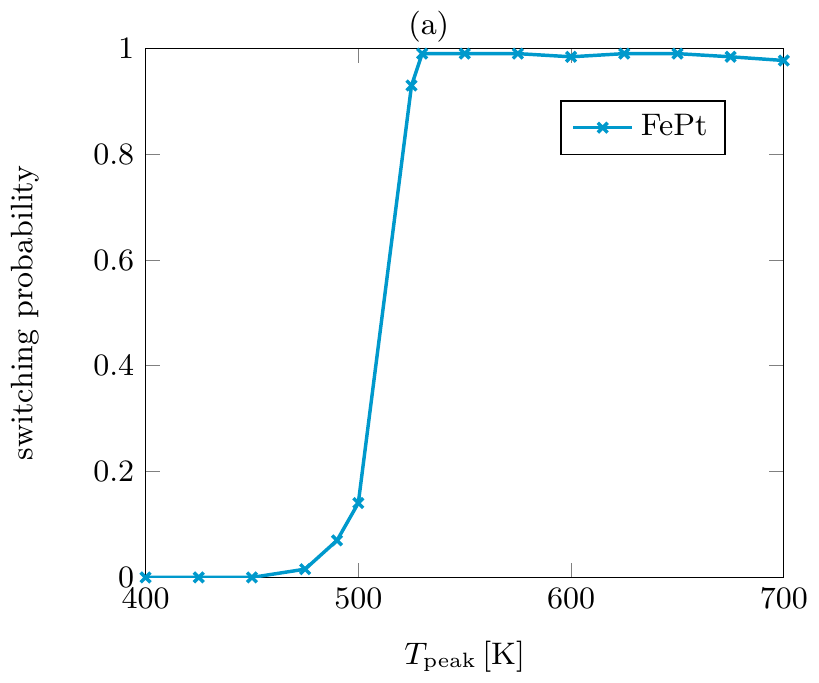}}
\subfigure{\includegraphics[width=\linewidth]{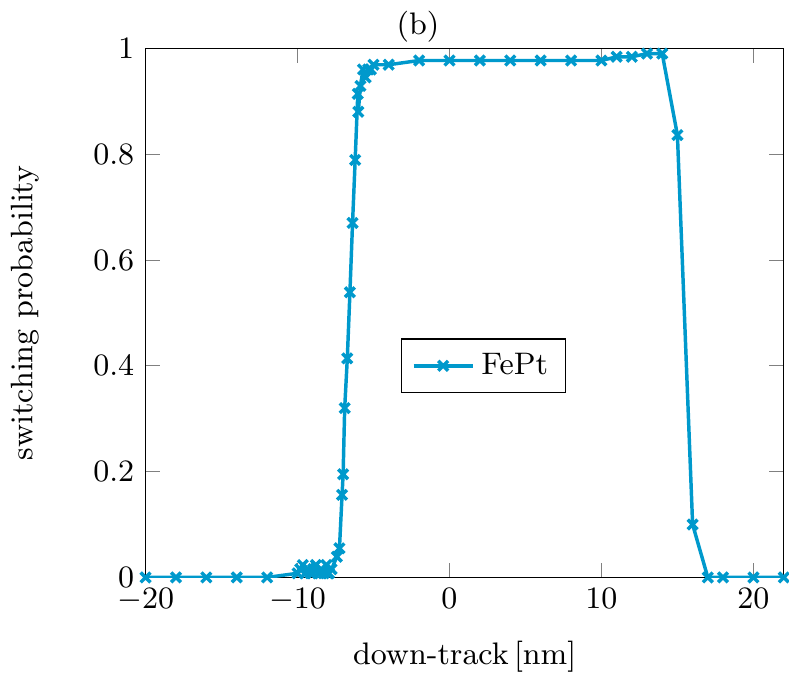}}
\caption{Switching probability curves of a FePt like hard magnetic grain. (a) $P(T_{\mathrm{peak}})$ which corresponds to off-track jitter for a fixed down-track position $x=0$\,nm. (b) Down-track jitter $P(x)$ for a fixed off-track position $y=0$\,nm and the peak temperature $T_{\mathrm{peak}}=700$\,K.}
\label{jitter}
\end{figure}

Both the off-track and the down-track jitter can be extracted from the switching probability phase diagram. 
The transition in off-track direction at a specific down-track position can be obtained by making a vertical cut in the phase diagram. For example, the off-track transition at down-track position $x=0$\,nm is marked by the dashed vertical line at $x=0$\,nm in \Cref{feptphase}.
On the other hand, the down-track transition is marked by a cut in the horizontal direction at a specific off-track position which is depicted by the corresponding peak temperature. For a peak temperature $T_{peak}=700$\,K, the switching probability curve at down-track position $x=0$\,nm and the transition curve in down-track direction at off-track position $y=0$\,nm ($T_{\mathrm{peak}}=T_{\mathrm{write}}$) can be seen in \Cref{jitter} (a) and (b).

\subsection{SOFT MAGNETIC LAYER}
 
\begin{figure}
\includegraphics[width=1.0\linewidth]{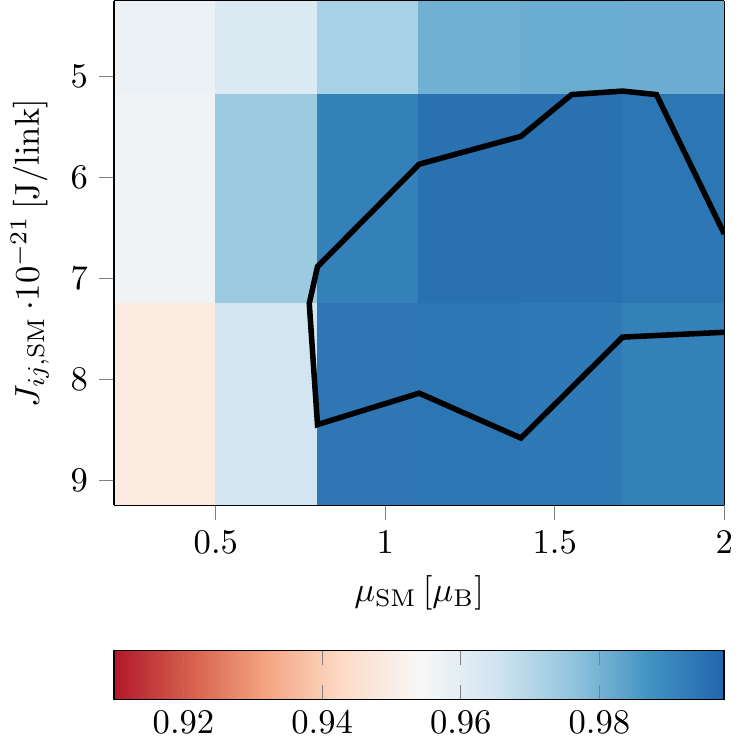}
  \caption{Phase diagram of a recording grain with $50\%$ hard magnetic material and $50\%$ soft magnetic material where the fitting parameter $P$ of \cref{distribution} can be seen for different configurations of the soft magnetic layer. The contour lines mark the areas with $P>0.992.$}
  \label{softsoft}
\end{figure}

An exchange coupled bilayer structure is considered. The parameters for a suitable soft magnetic composition are sought. Since the atomistic simulations are very time consuming, it is not possible to calculate a switching probability phase diagram for every material configuration. For this reason, only the switching probability curve along the off-track direction at $x=0$\,nm is calculated, again as a function of the peak temperature that corresponds to the respective off-track position. 
The result of these simulations is a switching probability curve $P(T_{\mathrm{peak}})$ for $400\,\mathrm{K} \le T_{\mathrm{peak}} \le 700\,\mathrm{K}$ which gives both, the maximum switching probability in the center of the grain and the off-track jitter. In a first optimization step, soft magnetic materials reaching complete switching in the center of the grains are pre-selected. 
To do this, the switching probability curves are fitted with a Gaussian cumulative distribution function
\begin{align}
\Phi_{\mu,\sigma^2}=\frac{1}{2} (1 + \mathrm{erf}(\frac{x-\mu}{\sqrt{2\sigma^2}}))\cdot P
\label{distribution}
\end{align}
with
\begin{align}
\mathrm{erf}(x)=\frac{2}{\sqrt{\pi}} \int_0^x e^{-\tau^2} d\tau,
\label{error}
\end{align}
where the mean value $\mu$, the standard deviation $\sigma$ and the mean maximum switching probability $P \in [0,1]$ are the fitting parameters. The standard deviation $\sigma$ determines the steepness of the transition function and is a measure for the transition jitter and thus for the achievable maximum areal grain density of a recording medium. The fitting parameter $P$ is a measure for the average switching probability for sufficiently high temperatures. 
In \Cref{softsoft}, one can see the fitting parameter $P$ for a recording grain as a function of different atomistic spin moments $\mu_{\mathrm{SM}}$ and different exchange energies $J_{ij,\mathrm{SM}}$ within the soft layer. 
Materials with $P$ less or equal 0.992 are not further considered. The phase diagram shows, that there are a few material compositions with sufficiently high $P$. For these materials, the fitting parameter $\sigma$ is additionally compared for the different configurations and the material with the lowest $\sigma$ is chosen. Two materials, namely that with $\mu_{\mathrm{SM}}=1.7\,\mu_{\mathrm{B}}$ and $J_{ij,\mathrm{SM}}=7.25\times 10^{-21}\,$J/link and that with $\mu_{\mathrm{SM}}=2.0\,\mu_{\mathrm{B}}$ and $J_{ij,\mathrm{SM}}=7.25\times 10^{-21}\,$J/link nearly have the same and the lowest $\sigma.$
Since the atomistic spin moment for the hard magnetic material is $\mu_{\mathrm{HM}}=1.7\,\mu_{\mathrm{B}}$ the same value is chosen for the soft magnetic material.  \\
In summary, the following parameters for the soft magnetic composition are chosen for further simulations: For the exchange constant within the soft magnetic material $J_{ij,\mathrm{SM}}=7.25 \times 10^{-21}$\,J/link is chosen. The exchange energy between the materials is set to $J_{ij}= \sqrt{J_{ij,\mathrm{HM}}\cdot J_{ij,\mathrm{SM}}}=6.13 \cdot 10^{-21}$\,J/link. The atomistic spin moment of the soft magnetic material is equal to that of the hard magnetic material namely $\mu_{\mathrm{SM}}=1.7\,\mu_{\mathrm{B}}$.

\subsection{BILAYER COMPOSITION}
\begin{figure}
\centering
\subfigure{\includegraphics[width=\linewidth]{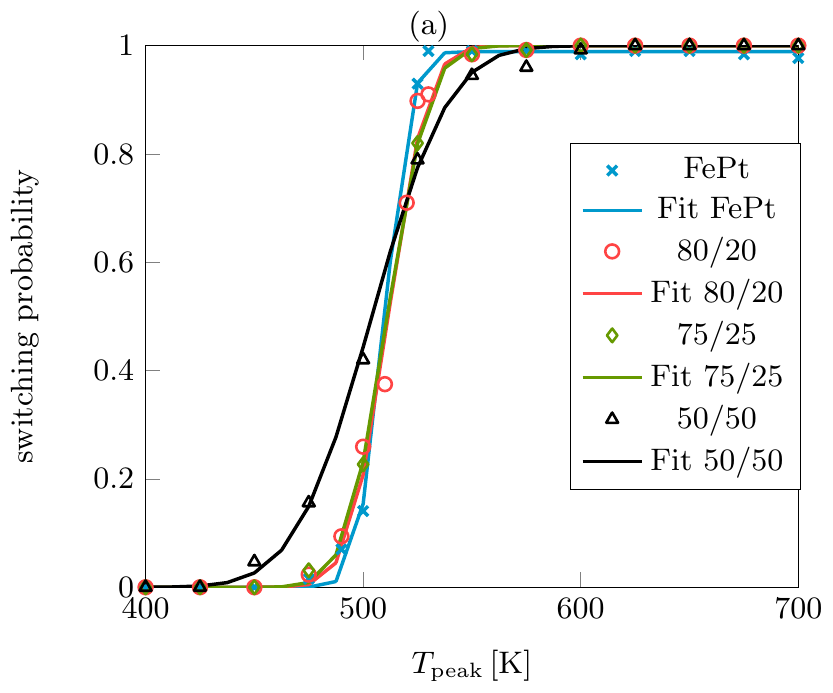}}
\subfigure{\includegraphics[width=\linewidth]{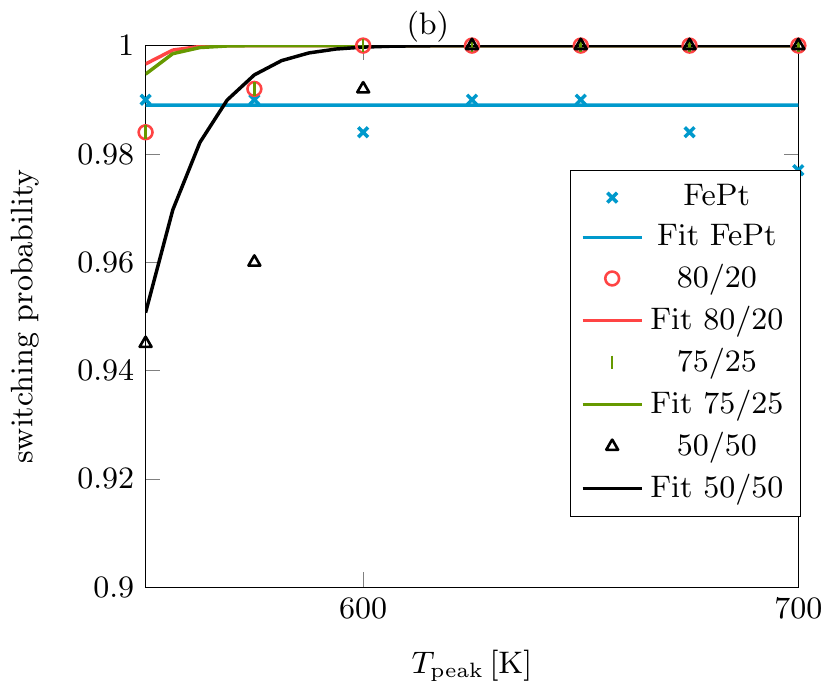}}
\caption{(a) Off-track jitter ($P(T_{\mathrm{peak}}$) curves) and corresponding fits of grains with different amounts of soft magnetic share at down-track position $x=0$\,nm for different peak temperatures $T_{\mathrm{peak}}$. (b) Zoomed transition and fitting curves for peak temperatures in a range from $550$\,K to $700$\,K.}
\label{transitionfit}
\end{figure}

\begin{center}
\begin{table}
\centering
\begin{tabular}{|l|c|c|r|}\hline
   Thickness HM$/$SM &  $T_{\mathrm{peak}}$ & $P_{\mathrm{switch}}$   \\
  \hline
	FePt & 654K  & $98.4\%$\\ 
  $90/10$ & 660K & $98.4\%$\\ 
  $89/11$ & 660K & $98.4\%$\\ 
  $88/12$& 666K& $99.2\%$\\
  $87/13$ & 666K& $99.2\%$\\
  $86/14$ &672K& $100\%$\\
  $85/15$ &672K& $100\%$\\
  $84/16$ &678K& $97.6\%$\\
  $83/17$ &678K& $97.6\%$\\  
  $82/18$ &678K& $97.6\%$\\
 $81/19$ &684K& $100\%$\\
  $80/20$ &690K&$100\%$\\
  $79/21$ &696K&$100\%$\\
  \hline
 \end{tabular}
\caption{Results for material compositions with different amounts of hard magnetic (HM) and soft magnetic (SM) share. The maximum switching probability is calculated at down-track postition $x=0$\,nm and for a peak temperature $T_{\mathrm{peak}}$ which is 20$\%$ higher than the Curie temperature of the material composition.}
\label{tablevariation}
\end{table}
\end{center}
With the chosen parameters, the amount of soft and hard magnetic material is optimized. The idea is to use as little soft magnetic material as possible to get narrow transitions, but as much soft magnetic material as necessary to get 100$\%$ switching probability in the simulations.
Phase diagrams are only computed for the most promising materials.
To find contemplable materials, again an off-track transition with switching probabilities at down-track position $x=0$\,nm and for temperatures in a range from 400\,K to 700\,K in steps of 25\,K is calculated for each material.\\
In \Cref{tablevariation} the resulting maximum switching probabilities for different material compositions with $T_{\mathrm{peak}}$ slightly above the Curie temperature of the material can be seen.
One observes that materials with an amount of soft magnetic material of 20$\%$ or more have a switching probability of 100$\%$ for temperatures around the Curie temperature. Thus, the transition curves of these materials are compared to that of FePt. This is done by fitting the transitions with a Gaussian cumulative distribution function as in \cref{distribution}. The important fitting parameter is the standard deviation $\sigma$ which is a measure for the transition jitter and thus for the achievable maximum areal grain density of a recording medium.
The transition curves and the corresponding fitting curves of the different materials can be seen in \Cref{transitionfit} for peak temperatures between $400$\,K and $700$\,K. 
Note, the fitting curve of FePt with $\sigma_{\mathrm{off},\mathrm{FePt}}=9.7$\,K shows that the switching probability does not reach 100$\%$ in the simulations.

\begin{figure}
\includegraphics[width=1.0\linewidth]{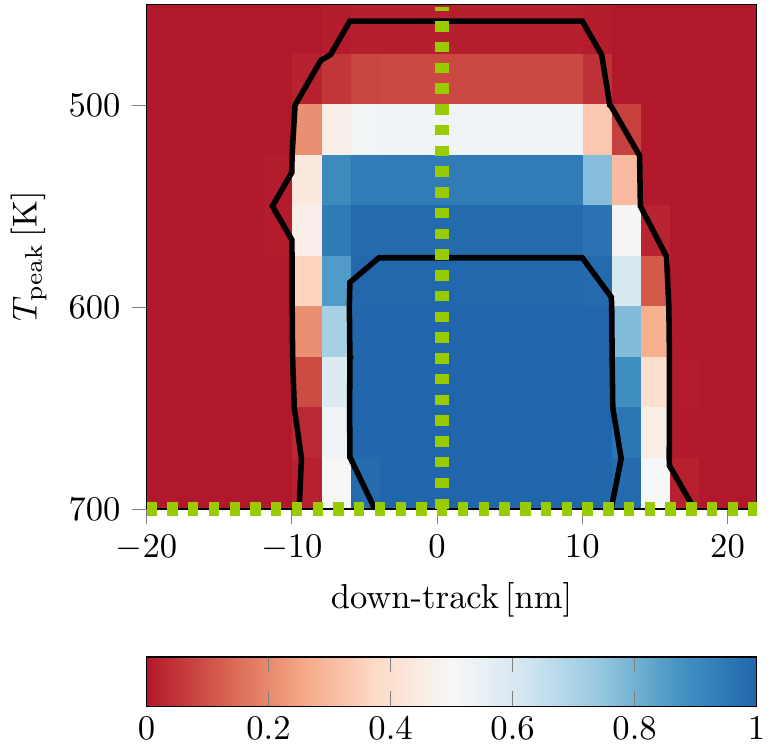}
  \caption{Switching probability phase diagram of recording grain consisting of a composition of 80$\%$ FePt like hard magnetic material and 20$\%$ Fe like soft magnetic material. 
The contour lines indicate the transition between areas with switching probability less than 1$\%$ (red) and areas with switching probability higher than 99.2$\%$ (blue). The dashed lines mark the switching probability curves in \Cref{transitionfit} and \Cref{downtrackjitterfepttest}.}
  \label{8020}
\end{figure}

For the other material configurations, full switching can be seen for sufficiently high temperatures. 
Further, in \Cref{transitionfit}, one can see that the off-track transition jitter gets larger for a higher amount of soft magnetic material. For 20$\%$ soft magnetic material, the transition is the steepest one among the bilayer compositions with $\sigma_{\mathrm{off},80/20}=14.2$\,K. Thus, it is much steeper than that of a material with 50$\%$ soft magnetic share for which $\sigma_{\mathrm{off},50/50}=27.87$\,K is almost twice as large as for 80/20. Actually, the transition of the composition with 20$\%$ soft magnetic material is the best compared to that of FePt although even here $\sigma_{\mathrm{off},80/20}$ is $46\%$ larger than $\sigma_{\mathrm{off},\mathrm{FePt}}$.\\
Since a grain with 80$\%$ hard magnetic material and 20$\%$ soft magnetic material (80/20) is the most promising material to fulfill our purpose, a switching probability phase diagram is calculated for 80/20. 
In \Cref{8020}, this switching probability phase diagram is illustrated. In contrast to the phase diagram of FePt (see \Cref{feptphase}), the 80/20 phase diagram shows complete switching also for higher peak temperatures. Indeed, the bilayer structure shows $100\%$ switching probability for peak temperatures higher than 550\,K in a range from down-track position $x=-10$\,nm to $x=6$\,nm. It can also be seen that the jitter in off-track and down-track direction of both materials does not differ much from the one of FePt.

\begin{figure}
\centering
\subfigure{\includegraphics[width=\linewidth]{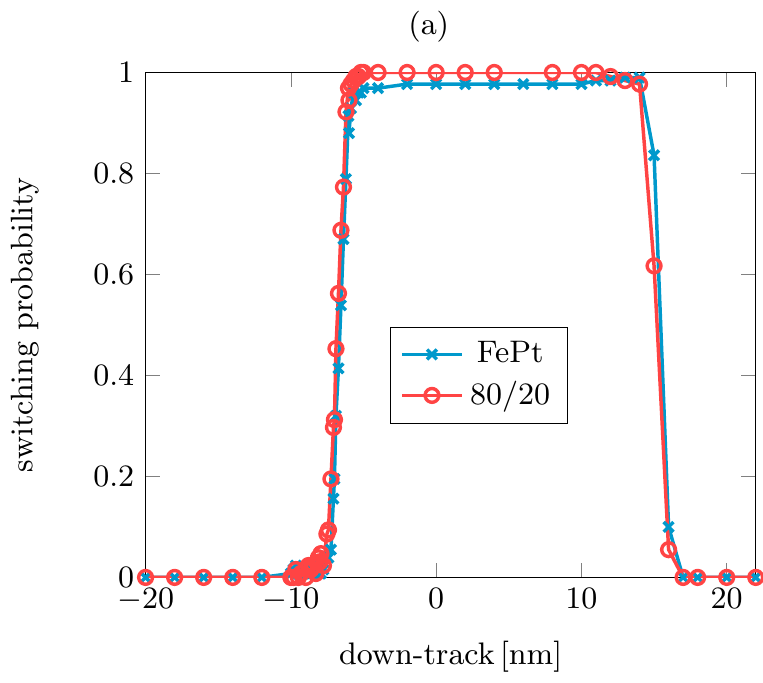}}
\subfigure{\includegraphics[width=\linewidth]{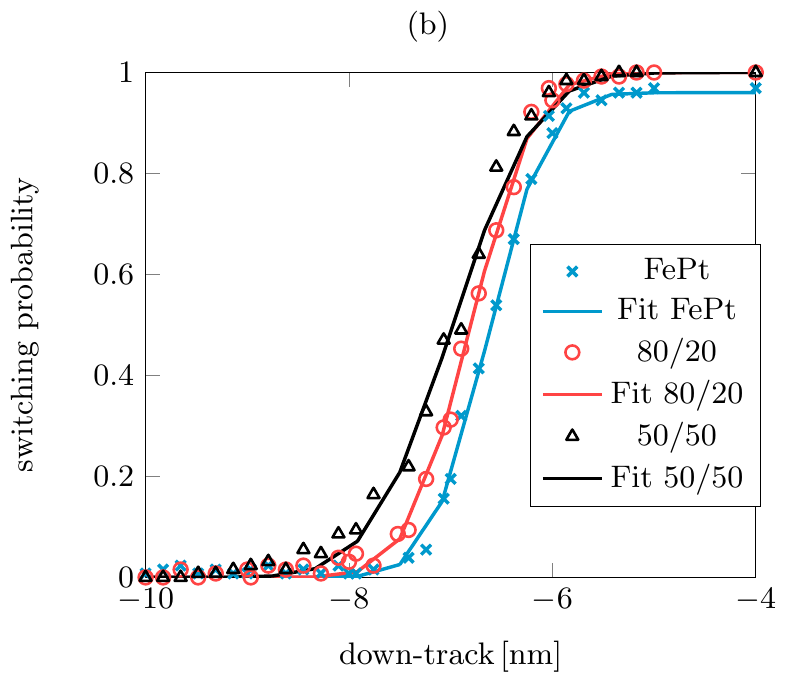}}
\caption{Comparison of down-track jitter at a peak temperature $T_{\mathrm{peak}}=700$\,K for a pure hard magnetic grain and a 80/20 hard/soft layer composition (a) for all down-track positions $x$ from  -\,20\,nm to 22\,nm. In (b) the jitter in down-track direction is shown more closely for down-track positions between $x=-10$\,nm and $x=-4$\,nm.}
\label{downtrackjitterfepttest}
\end{figure}

To compare the jitter of the materials more accurately, the jitter in down-track direction for one peak temperature and all down-track positions is calculated with higher resolution for both materials. Since writing of the grain starts at $T_{\mathrm{peak}}=475$\,K for both materials, the jitter is considered at the same peak temperature, $T_{\mathrm{peak}}=700$\,K, for both materials. The simulations are done for down-track positions $x$ between -20\,nm and 22\,nm with a resolution of $\Delta x=$2\,nm. Again, in the area of the transitions the resolution is finer, namely $\Delta x=0.16\dot 6$\,nm.  \\ 
In \Cref{downtrackjitterfepttest}, the results of these simulations can be seen. For $T_{\mathrm{peak}}=700$\,K, the 80/20 material reaches 100$\%$ switching probability whereas the switching probability of the pure FePt is clearly below 100$\%$. In contrast to the distinct switching probabilities of the materials, the transitions of FePt and 80/20 are almost the same. However, to compare the down-track jitter in more detail, the transitions are again fitted with a Gaussian cumulative distribution function like in \cref{distribution} and \cref{error}. For both temperatures the steepness of the transitions differs marginally. This can seen by the fitting values which are $\sigma_{\mathrm{down},FePt}=0.45$\,nm and $\sigma_{\mathrm{down},80/20}=0.49$\,nm. For comparison, a composition with 50$\%$ hard and $50\%$ soft magnetic material has a fitting value  $\sigma_{\mathrm{down},50/50}=0.64$\,nm.

\section{DISCUSSION}
To conclude, we simulated HAMR for a cylindrical recording grain ($d=5$\,nm, $h=10$\,nm) with an exchange coupled bilayer structure with graded Curie temperature. Here, the hard magnetic layer has a low Curie temperature and the soft magnetic layer a high Curie temperature. Our goal was to vary the composition and the amount of soft magnetic material such that at the same time both the AC noise and the DC noise are minimized. AC noise determines the distance between neighboring bits in bit-pat\-terned media. DC noise on the other hand limits the switching probability of a bit in bit-patterned media. Pure hard magnetic material shares high DC noise for a head velocity $v_h=20$\,m/s. In contrast, for a bilayer structure with too high soft magnetic fraction, the DC noise is significantly reduced but unfortunately the AC noise is significantly higher than for pure hard magnetic material.\\
Varying the soft magnetic composition showed that the atomistic spin moment does not influence the switching probability as much as the exchange interactions. Thus, a soft magnetic material with the same atomistic spin moment as the hard magnetic material and a higher Curie temperature was chosen. The used composition is similar to the one used by Vogler \textit{et al} \cite{areal}.
Further simulations to vary the amount of hard and soft magnetic material showed that more soft magnetic material leads to higher switching probabilities, whereas less soft magnetic material leads to narrower transitions. These results are not surprising. Because of the low coercivity of the soft magnetic layer and the exchange spring effect, the soft magnetic layer helps to switch the magnetization of the hard magnetic layer more reliably. Thus, the switching probability increases for a thicker soft magnetic layer. On the other hand, this increase of the switching probability is also visible for the jitter. Temperatures for which the grain does not switch for pure hard magnetic media start to switch for a bilayer composition. This explains the increase of the jitter in off-track and down-track direction for the bilayer structure.
We showed that a material composition consisting of 20$\%$ soft magnetic material and 80$\%$ hard magnetic material reduces both the AC and the DC noise.   \\
The 80/20 composition shows full switching in a wide range with a maximum switching probability $P_{\mathrm{switch}}>99.2\%$.  The transition jitter is comparable to that of FePt with the jitter parameters $\sigma_{\mathrm{off},80/20}=14.2$\,K and $\sigma_{\mathrm{down},80/20}=0.49$\,nm. These are only marginally different to those of FePt, i.e. $\sigma_{\mathrm{off},\mathrm{FePt}}=9.7$\,K and $\sigma_{\mathrm{down},\mathrm{FePt}}=0.45$\,nm. The 80/20 composition is much better than that of a 50/50 bilayer structure which has the jitter parameters $\sigma_{\mathrm{off},50/50}=27.87$\,K and $\sigma_{\mathrm{down},50/50}=0.64$\,nm.
Indeed, a $2$\,nm thick soft magnetic layer is in the expected range for optimal switching \cite{schrefl}, \cite{dieter}. The optimal thickness of the soft magnetic layer to reduce the switching field of the hard magnetic layer is around the exchange length between the soft and the hard magnetic layer \cite{thickness}. Since this exchange length is around $2$\,nm, our material fulfills this requirement perfectly. 

\section{ACKNOWLEDGEMENTS}
The authors would like to thank the Vienna Science and Technology Fund (WWTF) under grant No. MA14-044, the Advanced Storage Technology Consortium (ASTC), and the Austrian Science Fund (FWF) under grant No. I2214-N20 for financial support. The computational results presented have been achieved using the Vienna Scientific Cluster (VSC).

\nocite{*}
\bibliographystyle{unsrt}
\bibliography{noisehamr.bib}

\end{document}